\documentclass[twocolumn,superscriptaddress,floatfix,prl,showpacs]{revtex4-2}
\usepackage{graphicx}
\usepackage{amsmath}
\usepackage{amssymb}
\usepackage{color}
\usepackage{braket}
\usepackage{dsfont}
\usepackage{lineno}
\usepackage{float}

\begin{document}
\title{Quantum Phase Transition of Many Interacting Spins Coupled to a Bosonic Bath: static and dynamical properties}

\author{G. De Filippis}
\affiliation{SPIN-CNR and Dip. di Fisica - Universit\`a di Napoli Federico II - I-80126 Napoli, Italy}
\affiliation{INFN, Sezione di Napoli - Complesso Universitario di Monte S. Angelo - I-80126 Napoli, Italy}

\author{A. de Candia}
\affiliation{SPIN-CNR and Dip. di Fisica - Universit\`a di Napoli Federico II - I-80126 Napoli, Italy}
\affiliation{INFN, Sezione di Napoli - Complesso Universitario di Monte S. Angelo - I-80126 Napoli, Italy}

\author{A.~S.~Mishchenko}
\affiliation{RIKEN Center for Emergent Matter Science, Wako, Saitama, 351-0198, Japan}

\author{L.~M.~Cangemi}
\affiliation{SPIN-CNR and Dip. di Fisica - Universit\`a di Napoli Federico II - I-80126 Napoli, Italy}

\author{A. Nocera}
\affiliation{Department of Physics and Astronomy and Stewart Blusson Quantum Matter Institute, University of British Columbia, Vancouver, B.C., Canada, V6T 1Z1}

\author{P.~A.~Mishchenko}
\affiliation{NTT Secure Platform Laboratories, Tokyo 180-8585, Japan}

\author{M. Sassetti}
\affiliation{Dipartimento di Fisica, Universit\`a di Genova, I-16146 Genova, Italy}
\affiliation{SPIN-CNR, I-16146 Genova, Italy}

\author{R. Fazio}
\affiliation{SPIN-CNR and Dip. di Fisica - Universit\`a di Napoli Federico II - I-80126 Napoli, Italy}
\affiliation{ICTP, Strada Costiera 11, I-34151 Trieste, Italy}
\affiliation{NEST, Istituto Nanoscienze-CNR, I-56126 Pisa, Italy}

\author{N.~Nagaosa}
\affiliation{RIKEN Center for Emergent Matter Science, Wako, Saitama, 351-0198, Japan}
\affiliation{Department of Applied Physics, University of Tokyo, Tokyo 113-8656, Japan}

\author{V. Cataudella}
\affiliation{SPIN-CNR and Dip. di Fisica - Universit\`a di Napoli Federico II - I-80126 Napoli, Italy}
\affiliation{INFN, Sezione di Napoli - Complesso Universitario di Monte S. Angelo - I-80126 Napoli, Italy} 

\begin{abstract}
 By using worldline and diagrammatic quantum Monte Carlo techniques, matrix product state and a variational approach \`a la Feynman, we investigate the equilibrium properties and relaxation features of a quantum system of $N$ spins antiferromagnetically interacting with each other, with strength $J$, and coupled to a common bath of bosonic oscillators, with strength $\alpha$. We show that, in the Ohmic regime, a Beretzinski-Thouless-Kosterlitz quantum phase transition occurs. While for $J=0$ the critical value of $\alpha$ decreases asymptotically with $1/N$ by increasing $N$, for nonvanishing $J$ it turns out to be practically independent on $N$, allowing to identify a finite range of values of $\alpha$ where spin phase coherence is preserved also for large $N$. Then, by using matrix product state simulations, and the Mori formalism and the variational approach \`a la Feynman jointly, we unveil the features of the relaxation, that, in particular, exhibits a non monotonic dependence on the temperature reminiscent of the Kondo effect. For the observed quantum phase transition we also establish a criterion analogous to that of the metal-insulator transition in solids.   
\end{abstract}
\maketitle

Quantum phenomena play the most important role in quantum information technology, where information is stored, processed, and communicated following the laws of quantum physics \cite{gisin, nielsen}. Nowadays it is possible to develop quantum architectures, such as trapped ions \cite{duan, matsu}, superconducting qubits \cite{roch}, and Rydberg atoms \cite{saffman}, where quantum information applications can be implemented, exploiting the quantum mechanical features of many-body systems, i.e. coherence and entanglement. On the other hand, since no quantum system can be considered isolated from its environment, it is crucial to investigate the effects of decoherence, dissipation and entanglement induced by the rest of the universe, which limit the fidelity of the desired quantum operations.

The spin-boson model is the prototypical model of open quantum systems \cite{weiss}. It is the simplest realization of the Caldeira-Leggett model able to describe the quantum phase transition (QPT) from delocalized to localized states induced by the environment, and to shed light on the relaxation processes, in particular the dissipation and decoherence effects, in open quantum systems \cite{leggett,weiss,Hur,Breuer,revmod,sachdev}. The model consists of a two-level system, i.e. the elementary unit of a quantum computer, interacting with a set of quantum oscillators whose frequencies and coupling strengths obey specific distributions. Due to its versatility, it can capture the physics of a wide range of different physical systems going from defects in solids and quantum thermodynamics \cite{Lewis1988} to physical chemistry and biological systems \cite{rudnick,volker,huelga}. It has been also used to study trapped ions \cite{porras}, quantum emitters coupled to surface plasmons \cite{dzso}, quantum heat engines \cite{prx} or qubits strongly interacting with microwave resonators \cite{PhysRevA.97.052321}.

While the rich physics contained in the model involving a single qubit has been extensively addressed, only a limited set of works focus the attention on the characterization of QPT in the most interesting case of multiple two-level systems \cite{werner,werner1,orth,winter}. In particular, through Monte Carlo simulations \cite{winter}, it has been proved that a system of $N$ noninteracting spins coupled to a common bosonic bath undergo a QPT that is in the same class of universality of the single spin-boson model, i.e., in the Ohmic regime, a Beretzinski-Kosterlitz-Thouless (BKT) QPT occurs \cite{Kosterlitz_1973,kosterlitz1}. Furthermore the critical value of the coupling with the bath, $\alpha_c$, decreases asymptotically as $1/N$ with increasing $N$. At the heart of this result there is the {\em ferromagnetic} interaction among the spins induced by the bath. In the presence of an additional direct coupling among the spins the (thermo-)dynamics of the system may exhibit a much more complex behavior due to the competition between different interactions.

In this letter, beyond the coupling with the bosonic bath, we address also the effects of an antiferromagnetic interaction, with strength $J$, between all the $N$ spins, i.e. we investigate a frustrated model of $N$ spins coupled to a common bath. From the experimental point of view, current quantum annealing processors consist of manufactured interacting qubits \cite{prxlanting}. The aim is to determine which architecture, that is inevitably coupled to a thermal environment, is capable to preserve the quantum coherence. We prove that $J \ne 0$ is crucial to fulfill this objective. Indeed one of the main results of our work is that $\alpha_c$ goes to a constant by increasing $N$ at $J \neq 0$. Furthermore $\alpha_c$ is a monotonic increasing function of $J$. These results unveil a finite interval of $\alpha$ values, increasing with $J$, where the many qubit system is marginally influenced by the environment and then preserves quantum coherence even when $N$ is very large. By using matrix product state simulations (MPS) \cite{dmrg1,dmrg2, dmrg3, fishman, chinalex, snotesupp}, and combining the Mori formalism \cite{mori1} and a variational approach \`a la Feynman, we investigate also the relaxation processes. We not only confirm findings at the equilibrium by varying $\alpha$ at very low $T$, but observe also, at a fixed $\alpha$, a non monotonic behaviour with $T$ that is reminiscent of the Kondo effect \cite{Hewson}. Finally, by using the relaxation function, we establish, for the observed QPT, a criterion analogous to that of the metal-insulator transition in solids. Our proposal, addressing the changes of quantum Ising model in the presence of a tunable and common environment, can be experimentally realized in various open system quantum simulators \cite{review}, for instance, by extending the proposal based on the coupling between atomic dots and a superfluid Bose-Einstein condensate \cite{recati}. 

{\it The Model.} The Hamiltonian is written as:
\begin{equation}\label{eq:definitionH}
  H=H_{Q} + H_{B} + H_{I},
\end{equation}
where: 1) $H_{Q}=H_{\Delta}+H_{J}=-\frac{\Delta}{2}\sum_{i=1}^N \sigma_{x,i}+\frac{J \Delta}{4} \sum_{\substack{i,j=1,\\i < j}}^{N} \sigma_{z,i} \sigma_{z,j}$ describes the bare qubit contributions, $\Delta$ being the tunneling matrix element; 2) $H_{B}=\sum_i \omega_i a_i^\dagger a_i$ describes the bosonic bath; 3) $H_{I}= \sum_{j=1}^N \sigma_{z,j} \sum_i \lambda_i\left(a_i^\dagger+a_i\right)$ is the spin-bath interaction. In Eq.(\ref{eq:definitionH}), $\sigma_x$ and $\sigma_z$ are Pauli matrices with eigenvalues $1$ and $-1$. The couplings $\lambda_i$ are determined by the spectral function $F(\omega)=\sum_i\lambda_i^2\delta(\omega-\omega_i)=\frac{\alpha}{2}\omega_c^{1-s}\omega^s\Theta(\omega_c-\omega)$, where $\omega_c$ is a cutoff frequency. Here the adimensional parameter $\alpha$ measures the strength of the coupling and $s$ distinguishes the different kinds of dissipation. We focus our attention on the Ohmic regime ($s=1$), use units such that $\hbar=k_B=1$, and set $\omega_c=10 \Delta$.  

{\it Thermodynamic Equilibrium.} We investigate the physical features of this Hamiltonian by using three different approaches. The first of them is diagrammatic Monte Carlo (DMC) method, based on a stochastic sampling of the Feynman diagrams. It has been successfully applied to investigate polaron physics in different contexts \cite{mish5,mish1,mish2,mish3,mish4}. The second one is worldline Monte Carlo (WLMC) method, based on the path integrals. Here the elimination of the bath degrees of freedom leads to an effective Euclidean action \cite{bulla_1,weiss}:
\begin{equation}\label{eq:eqS}
  S=\frac{1}{2}\int_0^{\beta} d\tau \int_0^{\beta} d\tau^{\prime} \sum_{i,j} \sigma_{z,i}(\tau) K(\tau-\tau^{\prime}) \sigma_{z,j}(\tau^{\prime}),
\end{equation}
where $\beta=1/T$ ($T$ is the system temperature), and the kernel is expressed in terms of the spectral density $F(\omega)$ and the bath propagator: $K(\tau)=\int_0^{\infty} d\omega F(\omega) \frac{ \cosh \left[ \omega \left( \frac{\beta}{2}-\tau \right)\right]}{\sinh \left( \frac{\beta \omega}{2} \right)}$. In particular, for $\beta\rightarrow\infty$, the kernel has the following asymptotic behavior: $K(\tau)=\frac{\alpha}{2 \tau^2}$. The problem turns out to be equivalent to a classical system of spin variables distributed on $N$ chains (labelled by $i$ and $j$), each of them with length $\beta$, and ferromagnetically interacting with each other ($\tau$ and $\tau^{\prime}$ label the spins on the chains). The functional integral is done with Poissonian measure adopting a cluster algorithm \cite{rieger,bulla_1}, based on the Swendsen \& Wang approach \cite{swang}. This approach is exact from a numerical point of view and it is equivalent to the sum of all the Feynman diagrams. The third method is based on the variational principle, and, recently, has been successfully applied to the spin-boson model ($N=1$), where $\alpha_c \simeq 1$ \cite{giulio}. The idea is to introduce a model Hamiltonian, $H_{M}$, where one replaces the bath in Eq.(\ref{eq:definitionH}) with a discrete collection of fictitious modes, whose frequencies, $\tilde{\omega}_{i}$, and coupling strengths, $\tilde{\lambda}_{i}$, are variationally determined. A very limited number of these bosonic modes is enough to correctly describe, up to very low temperatures, any physical property, correctly predicting the QPT for any $s$ \cite{giulio}. 

\begin{figure}[H]
  \includegraphics[width=1.01\columnwidth]{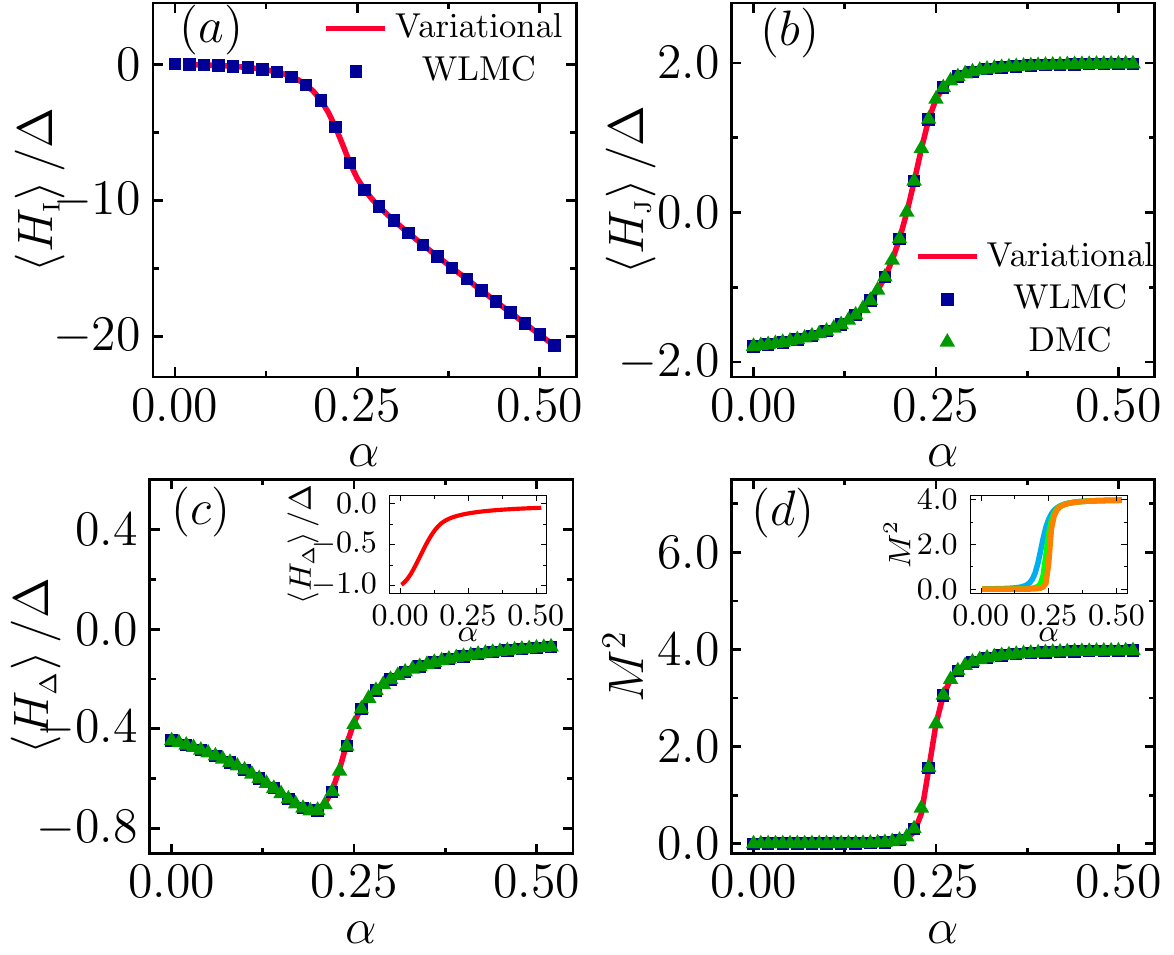}
  \caption{\label{fig:1} (color online)
   $\langle H_{I}\rangle$, $\langle H_{J} \rangle$, $\langle H_{\Delta}\rangle$, and $M^2$ vs $\alpha$ at $T=10^{-2} \Delta$, $N=2$ and $J=8$: comparison between DMC (triangles), WLMC (squares) methods and variational approach (solid line). Insets: c) $\langle H_{\Delta}\rangle$ vs $\alpha$ at $J=0$: d) $M^2$ vs $\alpha$ at: $T/\Delta=10^{-1}$ (blue), $10^{-2}$ (green), and $2 \times 10^{-3}$ (orange).}
\end{figure}

\begin{figure}[thb]
  \includegraphics[width=1.01\columnwidth]{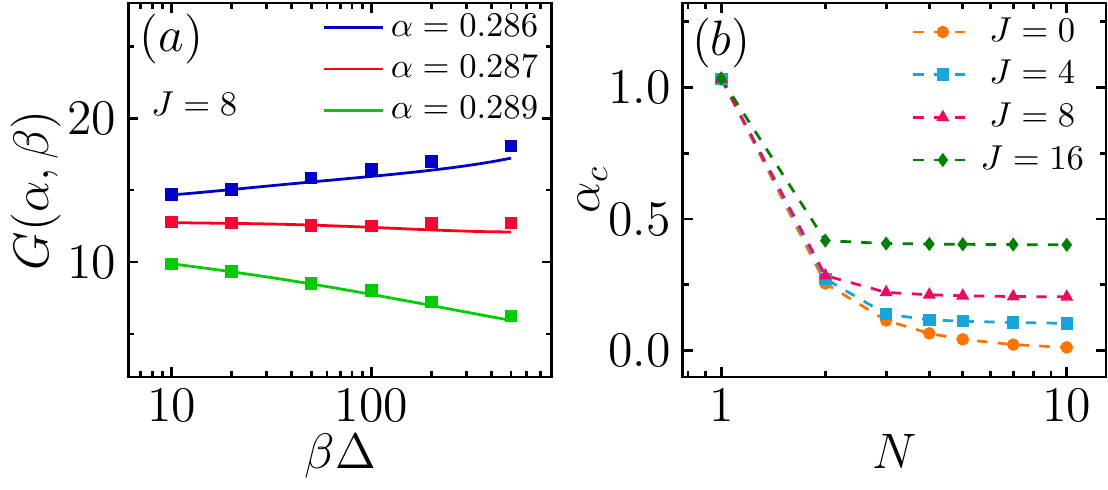}
  \caption{\label{fig:2} (color online)
    a) The function $G$ vs $\beta$ at $\alpha \simeq \alpha_c \simeq 0.287$ ($N=2$): WLMC (squares) vs variational approach (solid line); b) Phase diagram, $\alpha_c$ vs $N$, at different $J$.}
\end{figure}

In Fig.~\ref{fig:1} we plot $\langle H_{I}\rangle$, $\langle H_{J} \rangle$, $\langle H_{\Delta}\rangle$, and squared magnetization $M^2=\frac{1}{\beta}\int_{0}^{\beta} d\tau \langle S_z(\tau) S_z(0)\rangle$ as a function of $\alpha$, at $T=10^{-2} \Delta$ and $N=2$, with $S_z=\sum_{i=1}^N \sigma_{z,i}$. The plots point out the successful agreement between the $3$ approaches. As expected, by increasing $\alpha$: 1) the absolute value of $\langle H_I \rangle$ increases; 2) $\langle H_{J} \rangle$ increases in a monotonic way with a change of sign clearly indicating a progressive reduction of the effective antiferromagnetic interaction in favour of the ferromagnetic one; 3) $\langle H_{\Delta}\rangle$ shows a non monotonic behavior. The absolute value increases at weak coupling, where there is a different energy balance between $\langle H_{\Delta}\rangle$ and $\langle H_{J} \rangle$ with respect to $\alpha=0$: the spins tend to minimize $\langle H_{\Delta}\rangle$ at the expense of $\langle H_{J} \rangle$. On the other hand, by increasing $\alpha$ further, the effect of the dressing by the bosonic field prevails, inducing a progressive decrease of the effective tunnelling. Note that the non monotonic behavior is absent at $J=0$ (see inset). In this case, at $\alpha=0$, the average value of $H_{\Delta}$ is already minimized. 4) $M^2$ increases from $0$ to about $N^2$, in a steeper and steeper way by lowering $T$ (see inset), signaling an incipient QPT, that, independently on $N$, is again BKT QPT. Indeed, in a BKT transition, the quantity $M^2$ should exhibit a discontinuity at $\alpha_c$ and $T=0$ \cite{Kosterlitz_1973,kosterlitz1}. In order to get a precise estimation of $\alpha_c$, we adapt the approach suggested by Minnhagen et al. in the framework of the X-Y model \cite{minnhagen_1,minnhagen_2}.
In the present context, the roles of the chirality and the lattice size are played by squared magnetization and inverse temperature $\beta$, respectively. Defining the scaled order parameter $\Psi(\alpha,\beta)=\alpha M^2$, the BKT theory predicts: $\frac{\Psi(\alpha_c,\beta)}{\Psi_c}=1+\frac{1}{2(\ln\beta-\ln\beta_0)}$, where $\beta_0$ is the only fitting parameter and $\Psi_c=\Psi(\alpha_c,\beta\rightarrow\infty)$ is the universal jump that is expected to be equal to one. In this scenario, the function $G(\alpha,\beta)=\frac{1}{\Psi(\alpha,\beta)-1}-2\ln\beta$ should not show any dependence on $\beta$ at $\alpha=\alpha_c$. In Fig.~\ref{fig:2}a we plot the function $G(\alpha,\beta)$, as a function of $\beta$, for different values of $\alpha$. The plots clearly show that there is a value of $\alpha$ such that $G$ is independent on $\beta$. This determines $\alpha_c$. In Fig.~\ref{fig:2}b we plot the phase diagram of $\alpha_c$ vs $N$ for different $J$. While at $J=0$, $\alpha_c(J=0)$ decreases as a function of $N$, and asymptotically as $1/N$ by increasing $N$ \cite{winter}, for nonvanishing $J$ turns out to be rapidly independent on $N$. In order to explain this behavior we note that one has to take into account both the bare instantaneous antiferromagnetic coupling and the ferromagnetic interaction induced by the bath. The latter one includes both non-retarded contributions, with strength $\alpha \omega_c$ \cite{jnote}, and retarded contributions, that decrease as $1/\tau^2$ when $\beta\rightarrow\infty$ and give rise to BKT QPT. It occurs when $\alpha$ is greater than the maximum between $\alpha_c(J=0)$ and $\tilde{\alpha}$, $\tilde{\alpha}$ being the minimal value of $\alpha$ such that also the effective instantaneous interaction becomes ferromagnetic. It fulfills the relation $\frac{J \Delta}{4} - \tilde{\alpha} \omega_c=0$. Starting from $\alpha_c(J=0)$, $\alpha_c$ practically increases linearly with $J$. The more $J$ increases, the larger is the interval of $\alpha$ values with low environmental influences on quantum coherence.

{\it Relaxation towards Thermodynamic Equilibrium.} The relaxation function is the crucial physical quantity when the system is out of thermodynamic equilibrium. It represents the response of the system to a perturbation adiabatically applied from $t=-\infty$ and cut off at $t=0$, and can be calculated within the Mori formalism. It allows to reformulate, in an exact way, the Heisenberg equation of motion of any observable in terms of a generalized Langevin equation\cite{mori1}. Within this formalism, one introduces a Hilbert space of operators (whose invariant parts are set to be zero) where the inner product is defined by $(A,B)= \frac {1}{\beta} \int_{0}^{\beta} \left\langle e^{sH} A^{\dagger} e^{-sH} B \right\rangle ds$. Any dynamical variable $O$ obeys the equation:
\begin{eqnarray}
  \frac{dO}{dt}=- \int_{0}^{t} M_O(t-t{'}) O(t{'}) dt{'} + f(t),
  \label{moriO}
\end{eqnarray}
where the quantity $f(t)$ represents the ''random force'', that is, at any time, orthogonal to $O$ and is related to the memory function $M_O$ by the fluctuation-dissipation formula. The solution of this equation can be expressed as $O(t)=\Sigma_O(t) O+\tilde{O}(t)$, i.e. $\Sigma_O(t)=(O(t),O)/(O,O)$ describes the time evolution of the projection of $O(t)$ on the axis parallel to $O$ and represents the relaxation of the $O$ operator, whereas $\tilde{O}(t)$ is always orthogonal to $O$. We will focus our attention on $O=S_z$. If the system at $t=0$ is prepared at the thermal equilibrium in the presence of a small magnetic field $h$ along $z$ axis, by using the linear response theory \cite{kubo} and the Mori approach, it is possible to prove that $\Sigma_{z}(t)=\frac{\langle S_z(t) \rangle}{\langle S_z(0) \rangle}$, where $\langle S_z(t) \rangle$ is calculated in the absence of $h$. We proved that $\Sigma_{z}(z)$, the Laplace-transformed relaxation function, can be exactly expressed either as $\Sigma_{z}(z)=\frac{i}{z+iM_{z}(z)}$, i.e. \`a la Mori, or in terms of a weighted sum contributions associated to the exact eigenstates of the interacting system, each characterized by its own memory function:  
\begin{eqnarray}
  \Sigma_{z}(z)=\sum_n P_{n,z} \frac{i}{z + i M_{n,z}(z)},
  \label{sigman}
\end{eqnarray}
with $\sum_n P_{n,z}=1$ \cite{boltz, snotesupp}. In the case of the optical conductivity, where $O$ is current operator, this formulation resolves the difficulty to connect the Boltzmann transport theory and the Kubo formula.

\begin{figure}[H]
  \includegraphics[width=1.01\columnwidth]{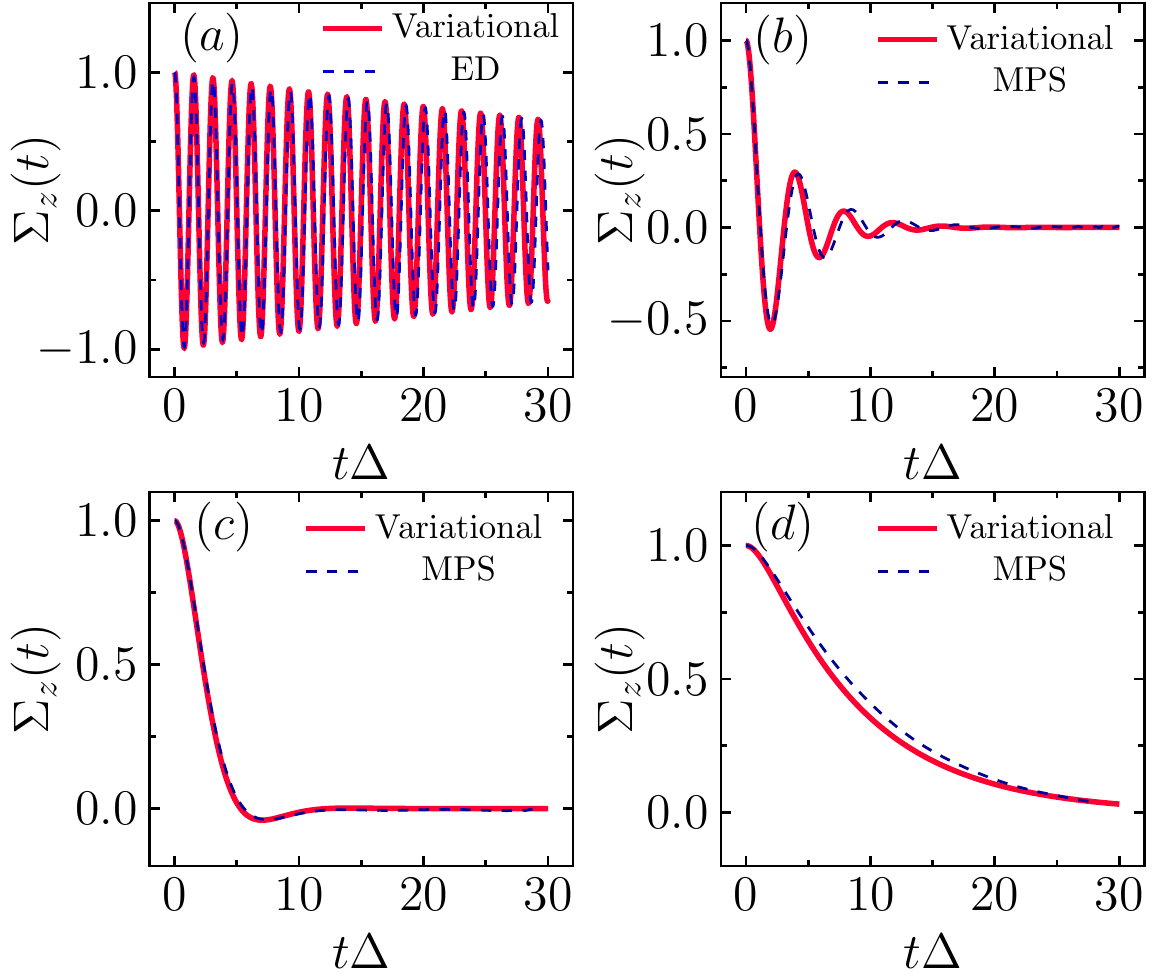}
  \caption{\label{fig:3} (color online)
    $\Sigma_z(t)$ at different values of $\alpha$ ($0.01$ (a), $0.15$ (b), $0.2$ (c), and $0.22$ (d)) at low $T$ and $J=8$: comparison between Feynman-Mori approach ($\beta \Delta=500$) and MPS and exact diagonalization methods ($T=0$) for $N=2$.}
\end{figure}

Here the current operator and the electric field are replaced by the spin operator and the magnetic field respectively, and $\Sigma_{z}(z)$ is the analogue of the optical conductivity, i.e. $\Sigma_{z}(z)=i \frac{(\chi(z)-\chi(z=0))}{M^2 \beta z}$, $\chi(z)$ being the magnetic susceptibility \cite{snotesupp}. It is straightforward to show that there is a relation between $\Sigma_{z}(z)$ and $\Sigma_{y}(z)$, i.e. between the two relaxation functions along $z$ and $y$ axes \cite{snotesupp}:
\begin{eqnarray}
  \Sigma_{z}(z)=\frac{i}{z}+\frac{(S_y,S_y)}{(S_z,S_z)} \Delta^2 \Sigma_{y}(z).
  \label{sigmayz}
\end{eqnarray}
Equation (\ref{sigmayz}) allows to define an effective gap: $\Delta_{eff}^2=\frac{(S_y,S_y)}{(S_z,S_z)} \Delta^2$. In particular it restores the bare gap $\Delta$ at $\alpha=0=J$. We emphasize that so far there isn't any approximation. Here, we combine, for the calculation of $\Sigma_{y}(z)$, the short-time approximation, typical of the memory function formalism \cite{mori1}, and the variational approach \`a la Feynman, by replacing the exact eigenstates of $H$ with that ones of the model Hamiltonian $H_M$ \cite{snotesupp}. These two approximations provide: $M_{n,y}(z)=\frac{i}{z} \Delta_n^2 + \frac{2}{\tau_n}$, i.e. it is possible to associate an effective gap, $\Delta_n$, and a relaxation time $\tau_n$ to any eigenstate of $H_M$. Here $M_{n,y}(z)$ is the memory function for $O=S_y$ (see Eq.(\ref{sigman})). We note also that, consistently, the following relation holds: $\sum_n P_{n,y} \frac{1}{\Delta_n^2}=\frac{1}{\Delta_{eff}^2}$. Equation (\ref{sigmayz}) allows us to obtain $\Sigma_{z}(t)$, the most important relaxation function:
\begin{eqnarray}
  \Sigma_{z}(t)=\Delta_{eff}^2 \sum_n P_{n,y} c_n(t),
  \label{sigmazt}
\end{eqnarray}
where
\begin{eqnarray}
  c_n(t)=\frac{1}{(\gamma_n^2+\frac{1}{\tau_n^2})} \left[ \cos(\gamma_n t)+\frac{1}{\gamma_n \tau_n}\sin(\gamma_n t) \right] e^{-\frac{t}{\tau_n}}
  \label{cn1}
\end{eqnarray}
if $\Delta_n^2 > \frac{1}{\tau_n^2}$, and
\begin{eqnarray}
  c_n(t)=\frac{\tau_n^{(+)}\tau_n^{(-)}}{2 \mu_n} \left[ \frac{1}{\tau_{n}^{(+)}} e^{-\frac{t}{\tau_{n}^{(-)}}}-\frac{1}{\tau_{n}^{(-)}} e^{-\frac{t}{\tau_{n}^{(+)}}} \right]
  \label{cn2}
\end{eqnarray}
if $\Delta_n^2 < \frac{1}{\tau_n^2}$. In Eq.(\ref{cn1}) $\gamma_n=\sqrt{\Delta_n^2 - \frac{1}{\tau_n^2}}$, and in Eq.(\ref{cn2}) $\mu_n=\sqrt{\frac{1}{\tau_n^2}-\Delta_n^2}$, $\frac{1}{\tau_{n}^{(+)}}=\frac{1}{\tau_{n}}+\mu_n$, and $\frac{1}{\tau_{n}^{(-)}}=\frac{1}{\tau_{n}}-\mu_n$. Independently on $N$, the predicted structure of $\Sigma_{z}(t)$ is always the same, i.e. a linear superposition of oscillating functions with decreasing amplitude and/or exponential functions. It is worth of mentioning that oscillation frequencies, $\gamma_n$, are determined by both $\Delta_n$ and $\tau_n$.

\begin{figure}[thb]
  \includegraphics[width=1.01\columnwidth]{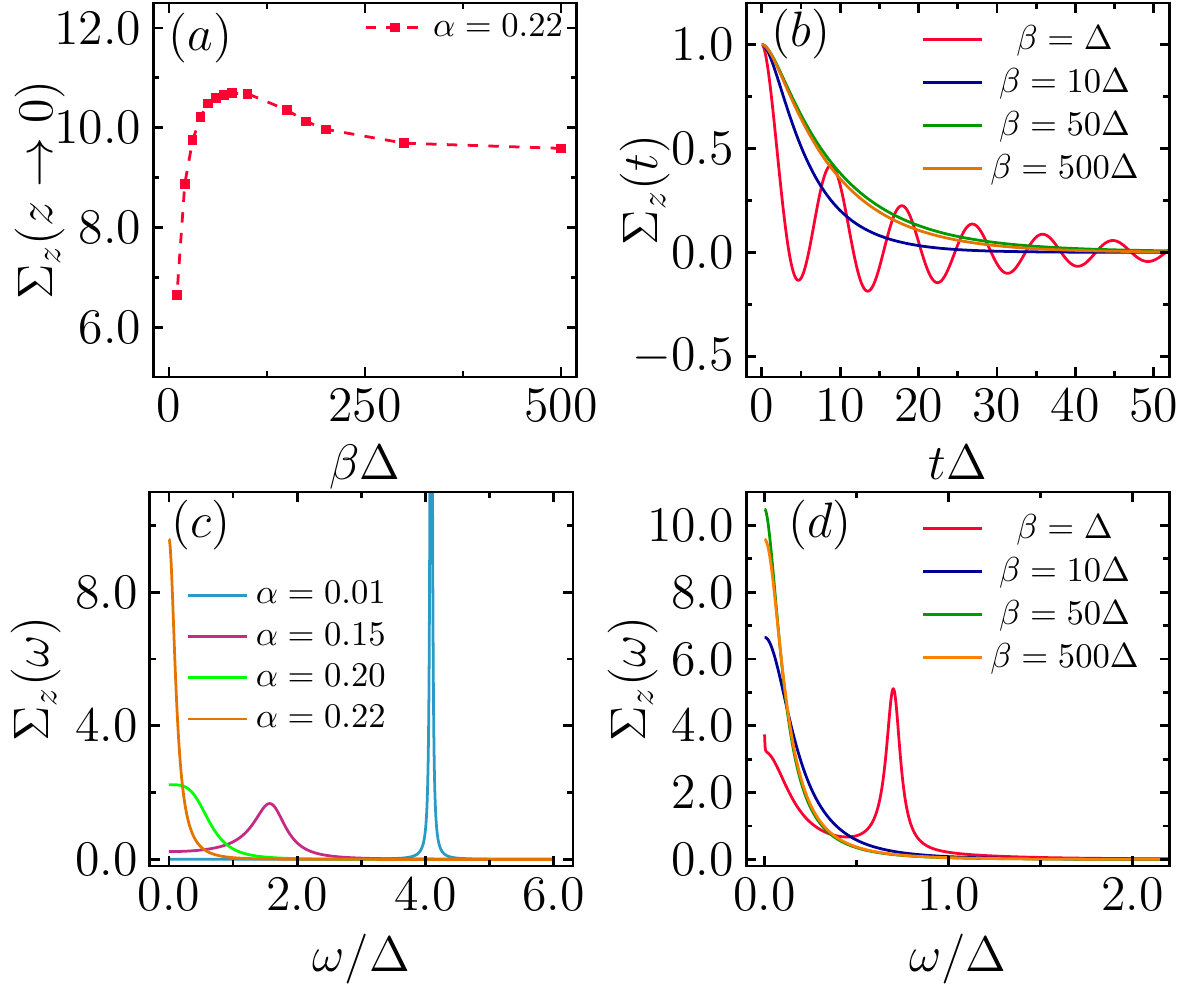}
  \caption{\label{fig:4} (color online)
    a) $\Sigma_z(z \rightarrow 0)$ vs $\beta$; b) $\Sigma_z(t)$ (in panel d) $\Sigma_z(z)$) at different $\beta$ and $\alpha=0.22$; c) $\Sigma_z(z)$ for different $\alpha$ ($\beta \Delta=500$). All the panels refer to $N=2$ and $J=8$.
  }
\end{figure}

In Fig.~\ref{fig:3} we plot the relaxation function at low T for different values of $\alpha$ at $N=2$ and $J=8$. The comparison with MPS and exact diagonalization methods \cite{exact} points out the effectiveness of our proposal at both short and long times and any spin-bath coupling. Firstly, by increasing $\alpha$, not only the amplitude but also the frequency of the oscillations reduces. When $\alpha$ is such that the quantity $\gamma_n$, corresponding to the ground state, becomes zero, i.e. $1/\tau_n=\Delta_n$, the relaxation becomes exponential. This is the analogue of the Toulouse point in the spin boson model with $N=1$ \cite{weiss}. By increasing $\alpha$ further, the relaxation time gets longer and longer, and, at $\alpha \ge \alpha_c$, the system does not relax, i.e. $\Sigma_{z}(t)=1$ independently on time $t$, signaling the occurence of QPT.

There are again two interesting observations. The first one regards the behavior of the quantity $\Sigma_z(z \rightarrow 0)$ as a function of $T$. It is the analogue of the conductivity in solids. Figure \ref{fig:4}a shows that this quantity has a non monotonic behavior with $T$, displaying a maximum at a finite temperature, that is reminiscent of the Kondo effect: it is the counterpart of the minimum (maximum) of the resistivity (conductivity) of the electron gas in the mapped model. The plots in Fig.~\ref{fig:4}b show that, correspondingly, also $\Sigma_z(t)$ exhibits a non monotonic behavior as a function of $T$. The second remarkable property is related to the possibility to introduce an alternative criterion to describe the QPT. Indeed the Fourier-transformed relaxation function obeys the following two sum rules: $\int_{-\infty}^{\infty} \Sigma_z(\omega) d\omega = \pi$ and $\int_{-\infty}^{\infty} \omega^2 \Sigma_z(\omega) d\omega= -\frac{4 \pi}{M^2 \beta}\langle H_{\Delta} \rangle$. On the other hand, while $\langle H_{\Delta} \rangle$ is a continuous function of the coupling $\alpha$ across QPT, the squared magnetization exhibits a discontinuity: $M^2 \beta$, when $\beta \rightarrow \infty$, tends to a finite constant depending on $\alpha$, for $\alpha < \alpha_c$, whereas, at $\alpha \ge \alpha_c$ diverges. It proves that, $\Sigma_z(\omega)$, at $T=0$ and $\alpha \ge \alpha_c$, becomes a $\delta$ function. QPT, in this model, exhibits the same characteristic of the metal-insulator transition in solids, provided that the optical conductivity is replaced by the spin relaxation function along $z$ axis. Figures \ref{fig:4}c and \ref{fig:4}d show the behavior of $\Sigma_z(\omega)$ at low $T$, as a function of $\alpha$, and, at a fixed value of $\alpha$, for different $T$, respectively. In Fig.~\ref{fig:4}c the peak of $\Sigma_z(\omega)$: first shifts from finite to zero frequency (in the time domain, the behavior of $\Sigma_z(t)$ from oscillating becomes exponential), then the peak width reduces and the maximum increases (at $T=0$ and $\alpha \ge \alpha_c$ a $\delta$ function is observed). On the other hand, Fig.~\ref{fig:4}d displays the non monotonic behavior in the frequency domain. By increasing $T$, first the peak width reduces, then increases and, at the same time, the maximum is located at a finite frequency.

{\it Conclusions.} We characterized QPT, static and dynamical features of $N$ spins antiferromagnetically interacting with each other and coupled to a common bath. We proved that, when $J \ne 0$, there is a finite range of values of $\alpha$ with low environmental influence on the spin phase coherence independently on $N$. We provided also an original way to address the spin relaxation processes, that exihibit a non monotonic behavior with $T$. Finally, for the observed QPT, we introduced a criterion analogous to that of the metal-insulator transition in solids.

\bibliography{main_text}{}

\begin{thebibliography}{52}
\bibitem [{jno()}]{jnote}%
  \BibitemOpen
  \href@noop {} {}\bibinfo {note} {When $H_{\Delta}=0$, i.e. in the absence of
  the tunneling, it is straightforward to show that the unitary transformation
  $e^S H e^{-S}$ with $S=\sum_{j=1}^N \sigma_{z,j} \sum_i (a_i^{\dagger}-a_i)
  \frac{\lambda_i}{\omega_i}$, allows to map the model on an Hamiltonian where
  the spin instantaneously interact with each other with strength $\frac{J
  \Delta}{4} - \alpha \omega_c$. Another way to estimate the strength of the
  instantaneous ferromagnetic spin-spin interaction consists in evaluating the
  Matsubara zero frequency of the Fourier series of the Kernel $K(\tau)$. We
  emphasize that BKT QPT occurs when $\beta \rightarrow \infty$, i.e. the
  length of the chains diverges and the effective ferromagnetic interaction
  induced by the bath approaches the form $\frac{1}{\tau^2}$. On the other
  hand, the bare direct interaction between the spins adds a positive
  contribution $\frac{J \Delta} {4}$ to the instantaneous ferromagnetic
  interaction, preventing BKT QPT when the total instantaneous coupling is not
  ferromagnetic, i.e. $\frac{J \Delta}{4} - \alpha \omega_c > 0$. Thus, to
  satisfy both requirements for BKT QPT, $\alpha_c$ must be larger than
  $\alpha_c(J=0,N)$ and $\tilde{\alpha} = \frac{J \Delta}{4 \omega_c} =
  \frac{J}{40}$.}\BibitemShut {Stop}%
\end{thebibliography}




%
%

\end{document}